\documentclass[vecphys]{svmult}

\usepackage{makeidx}         
\usepackage{graphicx}        
\usepackage{multicol}        
\usepackage[bottom]{footmisc}

\makeindex             

\begin{document}

\title*{Gas rich galaxies from the FIGGS survey}
\author{Jayaram N. Chengalur\inst{1}\and
Ayesha Begum\inst{2}\and
Igor D. Karachentsev\inst{3}\and 
Margrita Sharina\inst{3}\and
Serafim Kaisin\inst{3}
}
\institute{National Centre for Radio Astrophysics, TIFR, Pune University Campus, Ganeshkhind, Pune
\texttt{chengalur@ncra.tifr.res.in}
\and Institute of Astronomy, University of Cambridge, Madingley Road, Cambridge
\and Special Astrophysical Observatory, Nizhnii Arkhys 369167, Russia}
%
%
\authorrunning{Chengalur et al.} 
\maketitle

\section{Introduction}
\label{sec:1}

    The FIGGS (Faint Irregular Galaxy GMRT Survey) is aimed at creating a multi-wavelength
observational data base for a volume limited sample of the faintest gas rich galaxies. As
described in more detail in the contribution by Begum et al. in these proceedings, the
galaxies form  an HI flux and optical diameter limited subsample of the Karachentsev 
et al.(2004) catalog of galaxies within 10 Mpc.  The sample consists of  65 galaxies
with M$_B \widetilde{>}-14.5$ with median ${\rm{M_B}} \sim -13$ and a median HI mass 
$\sim 3 \times 10^7$~M$_\odot$. HI aperture synthesis data (from the Giant Meterwave
Radio Telescope - GMRT) has been obtained for all galaxies in the sample. Because
the GMRT has a hybrid configuration (see Swarup et al. (1991)) images at a variety
of spatial resolutions (ranging from $\sim 40^{``}$ to $\sim 3^{``}$) can be made
from a single GMRT observation run. Galaxies in the FIGGS survey have substantially 
lower M$_{\rm HI}$ and L$_{B}$ that typical of galaxies in earlier aperture synthesis
surveys.  The GMRT observations also used a velocity resolution ($\sim$1.6 kms$^{-1}$), 
that is $\sim$ 4 times better than most earlier interferometric studies of such 
faint dwarf galaxies. This high velocity resolution is crucial to detect large scale 
velocity gradients, which cannot be clearly distinguished in lower velocity
resolution observations (see e.g.  Begum et al. 2003a, 2003b, 2004a, 2004b, 
and for contrast Lo et al. 1993). In this paper we discuss two very gas rich
galaxies that were observed as part of the FIGGS survey, viz. NGC~3741 and AndIV.

\section{Two extremely gas rich galaxies}
\label{sec:twogals}

     NGC~3741 ($M_B \sim -13.13$) has M$_{\rm HI}$/L$_{\rm B} \sim 5.8$.
GMRT observations of this galaxy have been presented in Begum et al. (2005).
And IV, was originally thought to be  a satellite of the Andromeda (M31) galaxy. 
However, based on HST imaging Ferguson et al. (2000) argue that it 
is likely to be a background galaxy that happens to lie in projection close 
to the disk of M31. Consistent with this interpretation, they derive a 
distance of 6.11 Mpc for it  (using  the Tip of the Red Giant Branch technique)
-- this places And~IV  beyond the confines of the local group. The HI velocity 
measured  for this galaxy (i.e. 234 km/s; Braun et al. (2003)) is also substantially 
different from that of the nearest portion of the disk of M31. The galaxy has blue magnitude
of $\sim -12.37$, which implies that M$_{\rm HI}$/L$_{\rm B} \sim 13$.

\section{HI in NGC~3741 and And IV}

\begin{figure}
\centering \includegraphics[height=6cm]{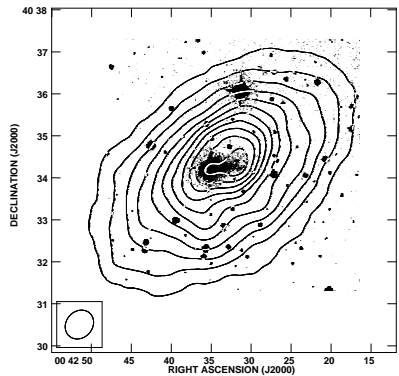}
\caption{ The GMRT moment 0 image of And IV  (at $\sim 44^{``}$ resolution), overlayed on the
          DSS image.}
\label{fig:jchengalur1}       
\end{figure}

   The GMRT observations (Begum et a. 2005) showed that NGC~3741 had an HI disk that extends to 
$\sim 8.3$ times its Holmberg radius. This makes it probably the most extended gas disk known. 
Our observations  allowed us to derive the rotation curve (which is flat in the outer regions) out to 
$\sim$38 optical scale lengths. NGC~3741 has a dynamical mass to light  ratio of $\sim$107 and 
is one of the "darkest'' irregular galaxies known. Follow up WSRT observations are presented
in Gentile et al. (2007).

\begin{figure}
\centering \includegraphics[height=6cm]{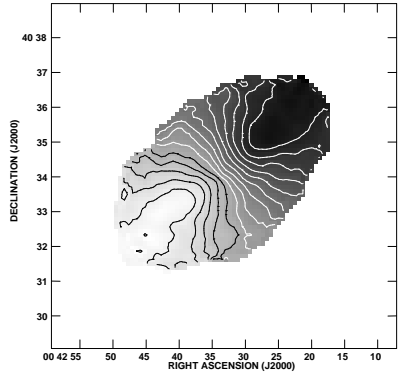}
\caption{ The GMRT moment 1 image of (velocity field) And IV  (at $\sim 26^{``}$ resolution).}
\label{fig:jchengalur2}       
\end{figure}

\begin{figure}
\centering \includegraphics[height=6cm]{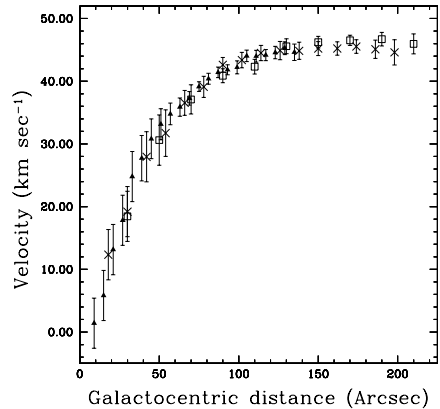}
\caption{ Rotation curve of AndIV as derived from the GMRT data.}
\label{fig:jchengalur3}       
\end{figure}

  For AndIV, the GMRT observations show that its gas disk extends out to $\sim 6$ Holmberg radii.
Fig.~\ref{fig:jchengalur1} shows the integrated HI emission from AndIV at 44$''\times 38''$ 
resolution,  overlayed on the digitised sky survey (DSS) image.  Fig.~\ref{fig:jchengalur2} 
shows the velocity field of AndIV at 26$^{''}\times 23^{''}$ resolution. The velocity field is 
regular and a large scale velocity gradient, consistent with systematic rotation, is seen across 
the galaxy. From the rotation curve Fig.~\ref{fig:jchengalur3} the ratio of the dynamical mass
to the blue luminosity is $M_{\rm dyn}/L_{B} \sim 237$!.

\begin{figure}
\centering \includegraphics[height=6cm]{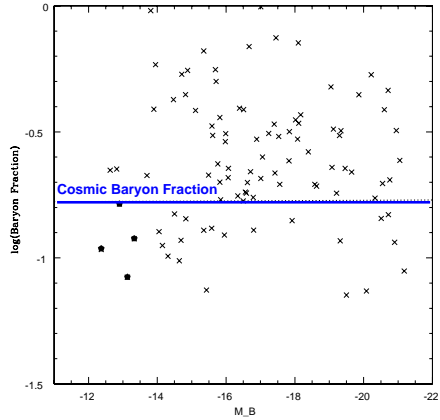}
\caption{ Baryon fraction (within the last measured point of the rotation curve) as a function
   of blue luminosity for a sample of galaxies with well measured rotation curves. The cosmic
   baryon fraction is shown as a horizontal line. Four gas rich galaxies, viz. DDO154, NGC3741,
   ESO~215~G?~009 (Warren et al. 2004) and And IV  are shown as solid points.}
\label{fig:jchengalur4}       
\end{figure}

     These very large dynamical mass to blue luminosity ratios naturally lead one to ask whether
extremely gas rich dwarf galaxies have abnormally small baryon fractions, i.e. have they just
been inefficient at forming stars, or did they end up with less than the typical baryon
fraction? The ratio of baryonic to dark matter is expected to systematically vary with
halo mass, since small halos are both inefficient at capturing hot baryons (for e.g. during
the epoch of reionization) and also because small halos are less able to prevent energy
input from star bursts from leading to escape of baryons (see e.g. Gnedin et al. 2002). In
Fig.~\ref{fig:jchengalur4} we show the baryon fraction (as determined at the last measured
point of the rotation curve) for a sample of galaxies with well measured rotation curves.
The average cosmic baryon fraction is shown as a horizontal line. As can be seen, there is a
large scatter in baryon fraction, and there is no systematic trend for a lower baryon fraction
in smaller galaxies. In particular although gas rich galaxies (shown as solid points in the figure),
have somewhat extreme baryon fractions, they lie within the range of that observed for galaxies
in general. As such these galaxies have got their ``fair share'' of baryons, but for some
reason have been unable to convert them into stars.

%
%
%
%
%

%
%



\printindex
\end{document}